\baselineskip=14pt
\hsize=6.0truein
\hfuzz=6pt
\vfuzz= 18pt
$ $
\vskip 1.4 in

\centerline{\bf Programming Pulse Driven Quantum Computers}

\vfil
\centerline{Seth Lloyd}
\vskip 1.0cm
\centerline{\it Complex Systems Group T-13}
\centerline{\it {\rm and} Center for Nonlinear Studies}
\centerline{\it Los Alamos National Laboratory}
\centerline{\it Los Alamos, New Mexico 87545}
\vskip 1.0cm
\noindent{\bf Abstract:} Arrays of weakly--coupled quantum systems
can be made to compute by subjecting them to a sequence of
electromagnetic pulses of well--defined frequency and length.  Such
pulsed arrays are true quantum computers: bits can be placed in
superpositions of 0 and 1, logical operations take place coherently,
and dissipation is required only for error correction.  
Programming such computers is accomplished by selecting the proper
sequence of pulses.
\vfil
\vskip 1.0in
\noindent{\bf Introduction}  

A recent paper proposed a technologically feasible quantum
computer.$^1$
This paper contains proofs of the results set forth in that paper.
The proposed computers are composed of 
arrays of weakly coupled quantum systems that are 
subjected to a sequence of 
electromagnetic pulses of well-defined frequency and length.
Selective driving of
resonances induces a parallel, cellular--automaton like logic on the
array, a method proposed by 
Teich {\it et al.}$^{2}$ for inducing a logic in arrays of quantum
dots.  In reference $1$, it is shown that this method extends to
arrays of quantum systems with generic weak couplings, and that the
resulting computers,  
when operated in a quantum--mechanically coherent
fashion, are examples of logically reversible computers
that dissipate less than $k_BT$ per
logical operation; dissipation is only required for error correction.$^{3-6}$
  In
fact, the systems are true
quantum computers in the sense of Deutsch:$^8$ 
bits can be placed in superpositions of 0 and 1,
quantum uncertainty can be used to generate random numbers, and
states can be created that exhibit purely quantum--mechanical
correlations.$^{7-12}$

In this paper, it is shown how such systems can be programmed. 
A simple sequence of pulses suffices to realize a universal parallel
computer.  The highly parallel operation of the system also allows
fast and robust error--correction routines.  A more complicated
sequence of pulses instructs the machine to perform arbitrary
unitary transformations on collections of quantum bits.

\bigskip
\noindent{\bf How it works}

For the purposes of
exposition, 
consider a heteropolymer, $ABCABCABC\ldots$, in which each unit
possesses an electron that has a long--lived
excited state.  For each unit, $A,B$ or $C$, 
call the ground state $0$, and the excited state
$1$.  Since the
excited states
are long--lived, the transition frequencies $\omega_A,
\omega_B$ and $\omega_C$ between the ground and
excited states are well-defined.  In the absence of any
interaction between the units, it is possible to drive transitions
between the ground state of a given unit, $B$ say, and the 
excited state by shining light at the resonant frequency $\omega_B$
on the polymer.$^{13-14}$
Let the light be in the form of a $\pi$ pulse, so
that $\hbar^{-1}
\int \vec \mu_B \cdot \hat e {\cal E}(t) dt = \pi$, where $\mu_B$ is
the induced dipole moment between the ground state and the excited
state, $\hat e$ is the polarization vector for the light that drives
the transition, and ${\cal E}(t)$ is the magnitude of the pulse
envelope at time $t$.  If the $\pi$ pulse is long compared with
$1/\omega_B$, so that its frequency is well--defined, and if the
polymer is oriented, so that each
induced dipole moment along the polymer has the same angle with
respect to $\hat e$, then its 
effect is to take each $B$ that is in the ground state
and put it in the excited state, and to take each $B$ in the
excited state and put it in the ground state.

Now suppose that there are local interactions between the units of
the polymer, given by interaction Hamiltonians $H_{AB}, H_{BC},
H_{CA}$.  
Almost any local interaction will do.
Consider first the case in which these interaction
Hamiltonians are diagonal in the original energy eigenstates for
each unit (the effect of off--diagonal terms is considered below).
The only effect of such interactions is
to shift the energy levels of each unit as a function of the energy
levels of its neighbors, so that the resonant frequency $\omega_B$,
for instance, takes on a value $\omega^B_{01}$ if the
$A$ on its left is in its ground state and the $C$ on its right is
in its first excited state.  If the resonant frequencies for all
transitions are 
different for different values of a unit's neighbors, then the
transitions can be driven selectively: if a $\pi$ pulse with
frequency $\omega^B_{01}$ is applied to the
polymer, then all the 
the $B$'s with an $A=0$ on the left and a $C=1$
on the right  
will switch from 0 to 1 and from 1
to 0 are.  If all transition frequencies are different, these are the
only units that will switch.  Each
unit that undergoes a transition coherently emits or absorbs a
photon of the given frequency: no dissipation takes place in the
switching process.

Driving transitions selectively by the use of resonant ${\pi}$ pulses
induces a parallel logic on the states of the polymer: a particular
resonant pulse updates the states of all units of a given type
as a function of its previous state and the states of its
neighbours.  All units of the given type with the same values for
their neighbours are updated in the same way.
That is, applying a resonant pulse to the polymer effects the action
of a cellular automaton rule on the states of units of the polymer.$^{2,15}$
The cellular automaton rule is of a particularly simple type: if the
neighbours of the unit to be switched take on a specific pair of
values, then a permutation that interchanges two states of that unit
is induced.  Since an arbitrary permutation of $N$ states can be
built up of successive interchanges of two states,
one can by the proper sequence of pulses realize 
{\it any} cellular automaton rule that permutes first the states of
one type of unit as a function of its neighbours, 
then permutes the states of another type of unit, 
then another, etc.  Any reversible cellular automaton rule in which
updating takes place by acting
first  on one type of unit, then another, then another, must be of
this form, a permutation $\pi^X_{ij}$ that induces the permutation
$\pi_{ij}$ on the states of all units of type $X$ for all different
states $ij$ of the neighbours of
a unit.   

The system is much more computationally powerful than a simple
cellular automaton, since one can change the cellular automaton rule
from step to step by changing the sequence of pulses
applied.  
Wolfram$^{15}$ discussed variable rule cellular automata
in the form of coupled `Master--Slave' automata, in which the state
of the Master cellular automaton varies the rule of the Slave
automaton.  The systems discussed here are more general than
Wolfram's example, however, for the simple reason that the `Master'
is the programmer of the computer: any sequence of pulses, any
program, is allowed.  In fact, as will be shown below,  
by selecting the sequence of pulses, one can make even the
simplest of such 
systems perform any computation that one desires: pulse driven
quantum computers are universal digital computers.  

It is worth noting that any such variable rule cellular automaton
with $M$ types of
unit, and $m_i$ states for the $i$-th unit, is equivalent in terms
of its logical operation to any other such system with the same $M$
and $m_i$.   
In the following exposition of sequences of pulses needed to program
such variable rule cellular automata, we will concentrate on
automata that are one--dimensional, have two states per site, and
have three different types of units, $A,B,C$, as above.  We will
show that even such extremely simple systems can be made to perform
arbitrary computations.  One--dimensional systems with more states
per site, or more types of units, or both, are then also
computationally universal.  We will also prove computational
universality for a
one--dimensional system with only two types of units, $A,B$, in
which $B$ has three states, one of which exhibits a fast decay to
the ground state.  Whether a two--unit, two state reversible
variable rule automaton is computationally universal is an open
question.  Although the exposition here will concentrate on
one--dimensional systems, we will also note explicitly when the
techniques supplied can be generalized to systems of higher
dimension.

\bigskip
\noindent{\bf Loading and unloading information}

A simple sequence of pulses allows one to load information onto
the polymer.  
There is one unit on the polymer that can be controlled independently
--- the unit on the end.$^{2}$  
Simply by virtue of having
only one neighbour, the unit on the end in general has different resonant
frequencies from all other units of the same type.  
Suppose this
unit is an $A$: the resonant frequencies $\omega^{A:end}_i$
for this unit are functions only of the state $i$ of the $B$ on its
right.  If these resonant frequences are different from the resonant
frequencies $
\omega^A_{ij}$ of the $A$'s in the interior of the polymer, then one
can switch the end unit from $0$ to $1$ on its own.

Suppose that 
all units are initially in
their ground state.
To load a $1$ onto the polymer, apply a $\pi$ pulse at frequency
$\omega^{A:end}_0$.  This pulse switches the end
unit to $1$.  To move this $1$ along the polymer, apply a $\pi$
pulse with frequency $\omega^B_{10}$.  The only $B$ that responds to
this pulse is the first: it will switch to $1$.  Now apply a pulse
with frequency $\omega^{A:end}_1$.  This pulse
switches the $A$ on the end back to $0$.  (This act of reversibly
restoring a bit to zero using a copy of the bit is called
`uncopying,' and is typical of reversible computation
schemes.$^{3-6}$)  

To load an arbitrary sequence onto the computer,   
note first that a sequence of $\pi$ pulses with frequencies
$\omega^B_{10}$, $\omega^B_{11}$, $\omega^A_{01}$, $\omega^A_{11}$,
$\omega^B_{10}$, $\omega^B_{11}$, 
swaps information between adjacent
$A$'s
and $B$'s, taking whatever information is
registered in each $A$ (except the $A$ on the end)
and exchanging it for the information in its
neighbouring $B$.  Adding in the middle of this sequence
a pulse with frequency
$\omega^{A:end}_1$ causes the $A$ on the end to swap information
with the $B$ on its right as well. 
Similarly, one can exchange information between
adjacent
$B$'s and $C$'s and $C$'s and $A's$.  (No additional pulses to
address the end unit is required for these exchanges.) 

To load an arbitrary sequence
of values, $a_1b_1c_1\ldots a_nb_nc_n$ onto the polymer, first
load $b_n$ onto the $A$ on the end.  Swap information between $A$'s
and $B$'s, including the $A$ on the end.  
Now load $a_n$ onto the $A$ on the
end.  Swapping the information in the $B$'s with the information in
the $C$'s, then the $A$'s with the $B$'s, 
then the $C$'s with the
$A$'s, then the $B$'s with the $C$'s moves the bit $b_n$ to the
first $A$ from the end, and $a_n$ to the first $C$ from the end.
Now load $b_{n-1}$ onto the $A$ on the end.  Swap information
between $A$'s and $B$'s, etc.  Continuing this process loads
$a_1b_10a_2b_20\ldots a_nb_n0$ onto the first $n$ $ABC$'s.  

To load on the $c$'s as well, note that the sequence of swaps, $B$'s
with $C$'s, $A$'s with $B$'s,
has the effect of taking the information on the $A$'s and $B$'s, and
moving each bit one unit to the right, while taking the information
on the  
$C$'s, and moving each bit two units to the left.  Continuing with
the sequence of swaps, $C$'s with $A$'s, $B$'s with $C$'s, then
$A$'s with $B$'s, $C$'s with $A$'s, moves each $ab$ pair to the $AB$
one triple, $ABC$ to the right, while moving each $C$ two triples to
the left.  The same set of swaps in opposite order ondoes the
motion, moving each $ab$ one triple to the left, and each $c$ two
triples to the right.  It is clear that one can, by the proper
sequence of swaps, shift the information in one type of units by any
amount with respect to the information in the other types of units
(subject to the constraint that the overall `center of gravity' of
the array remains fixed, in the sense that sums of the displacements
of the information in all types of units remains zero). 
To load on the $c$'s, move the string $a_1b_1\ldots a_nb_n$ $2n$
units to the right, then add on the $c$'s at intervals of 3 units,
starting with $c_n$, then with $c_{n-1}$, etc.,
moving the $a$'s and $b$'s left by one unit for each shift of the
$c$'s right by 2 units.  When all the $c$'s have been loaded, they
will be paired with the proper $a$'s and $b$'s in the first $n$
triples.  

(It is clear from the discussion of loading information above 
that for an array with $M$ different types of units, it is simpler to
load information in chunks whose size does not
exceed $M-1$ bits, since $M-1$ bits can be
translated as a block.  Whenever possible, we will use schemes for
computation that require chunks of no greater than this size.  This
practice will prove important when error--correction is introduced.)

This technique clearly works when there are two or more different
types of units.  The different types of units can have different numbers
states, although the maximum amount of information that can be
stored and transferred per unit is limited by the type of unit
with the smallest number of
states.
For arrays of more than one dimension, note that the same 
type of unit will tend to have distinct
resonant frequencies if it is on a corner, edge, face, or in the
interior of the array.  In addition, two of the same type of unit on
two different corners (edges, faces) will tend to have distinct resonant
frequencies if the type and configuration of their neighbours are
different.  To load an arbitrary block of bits onto a
multi--dimensional array, one starts at a corner, loads an arbitrary
string onto an edge, moves it inward one unit onto a face, loads in
the next string, moves the two strings a unit further onto the face,
and continues until the face contains the first 2-d cross--section
of the block.  This cross section can be moved into the interior of
the block location while the next 2-d section is built up.  Etc.
Symmetries of the array can interfere with this process.

\bigskip
\noindent{\it Unloading information}

There are several ways to get information off the polymer.  All
involve a certain amount of redundancy, since detection efficiencies
for single photons are not very good.  The simplest way is to have
many copies of the polymer.  The same sequence of pulses will induce
the same sequence of bits on each copy.  To read a bit, one applies
a sequence of pulses that moves it to the end.  Then one applies two
$\pi$ pulses, with frequencies $\omega^{A:end}_{0,1}$.
If either of these pulses is attenuated, then the bit on the end is
a $1$; if either is amplified, then the bit is a $0$.  
This method has the
disadvantage that all bits must be moved to the end of the polymer
to be read.

If the light in the $\pi$ pulses can be focussed to within a radius
of a few wavelengths, information can also be read out in parallel,
simply by copying the bit that is to be read out onto all or most
units of the same type within a few wavelength neighbourhood, and
then seeing whether $\pi$ pulses aimed at that neighbourhood are
attenuated or amplified.  (The error--correction schemes described
below already require some redundancy.)
Other schemes that require less
redundancy exist.  For example, if the end unit has a fast decay mode
(as described below in the section on dissipation), the signal for a
bit being a 0 or 1 can be a photon of a different frequency than
that of the switching pulse.  Only a 
small numbers of such photons in a
distinct frequency channel need be present to be detected with high accuracy. 

\bigskip
\noindent{\bf Computation}

Once information is loaded onto the polymer, a wide variety of
schemes can be used to process it in a useful fashion.  
It is not difficult to find sequences of pulses that 
realize members of the following class of
parallel processing computers.

The polymer is divided up into sections of equal length.
By choosing the proper sequence of pulses, and by properly
formatting the input information, one can simulate the
action of any desired reversible logic circuit on the information
within each section.  (Since every logical action described up until
now is reversible, the entire circuit must be reversible: the logical 
operation induced by a sequence of $\pi$ pulses
can be reversed simply by applying the same sequence in reverse order.)
The logic circuit realized is, of course, the
same for each section, although the initial information on which the
circuit acts can be different from section to section.  A second
sequence of pulses allows each section to exchange an arbitrary
number of bits with the sections to its left and right.  Input and
output can be obtained from the sections on the end, as above, or
from each individual section using focussed light.  

By choosing the proper section size and sequence of pulses, one can
then realize a string of identical microprocessors of arbitrary
reversible
circuitry, each
communicating with its neighbours.  Such a device is obviously
computationally universal, in the sense that one can embed in it the
operation of a reversible universal Turing machine.  A device with the parallel
architecture described here, however, is likely to be considerably
more useful than a Turing machine
for performing actual computations.  
The number of pulses required to realize
such a machine is proportional to the length of the wires, measured
in terms of the number of units over which bits must be transported,
and number
of logic gates in one microprocessor.

We justify the above assertions by giving a recipe for constructing
a sequence of pulses that realizes the parallel computer described.

The sequence of $\pi$ pulses with frequencies,
$\omega^C_{10}$, $\omega^C_{11}$, $\omega^B_{11}$,
$\omega^C_{10}$, $\omega^C_{11}$, induces the operation of a Fredkin
gate on each triple $ABC$: a Fredkin gate is a 
binary gate with three inputs, $X,Y,Z$ and three outputs $X',Y',Z'$
in which $X'=X$, and $Y'=Y$, $Z'=Z$ if $X=0$; $Y'=Z$, $Z'=Y$ if
$X=1$.$^4$  (Note that this sequence is closely related to, but
simpler than, the  
set of pulses required to exchange
information between the $B$'s and the $C$'s:
the only difference is the lack of
the pulse with frequency $\omega^B_{01}$).
That is, if $X=0$, all three inputs go through unchanged; if
$X=1$, the second and third input are exchanged: a Fredkin gate
effects an exchange of information between two units conditioned on
the value of a third.  Fredkin gates
suffice to give the logical operations $AND$, $OR$, $NOT$ and
$FANOUT$, which in turn form a basis for digital computation.  

These two operations, exchange of information between adjacent units,
and conditional exchange of information, suffice to create the sort
of parallel computer described.  The trick is encoding the
information in the proper format.   Any reversible logic circuit can
be constructed from Fredkin gates that operate first on a given triplet
of bits, then on another triplet of bits, etc.  But the sequence of
pulses
described in the previous paragraph applies a Fredkin gate to all
collections of three bits at once.  This extreme parallelism can be
overcome as follows.  

Each of the processors in the parallel design described above can be
described by the same logical circuit.  Consider an implementation
of this circuit by Fredkin gates: the circuit design consists of
wires that move bits to the proper location, and Fredkin gates that
then operate on the the proper bits three by three.  There are many
possible ways to deliver a sequence of pulses that causes the
computer to realize the operation of the desired logic circuit.
Here we consider a few of the simplest.

Suppose that the implementation of the circuit in terms of Fredkin
gates requires $N$ bits of input, or in the circuit diagram, $N$
wires leading into the circuit.  The $k$th processor will be
realized on the $2kN$ to $2(k+1)N$ triples $ABC$.
Let the $N$ bits $x_1,\ldots, x_N$ that are to be
input into the $k$th processor be loaded onto the first $N$ $A$ units
of this section, and  
let the $N+1$st triple $ABC$ contain $011$.  Let all the remaining
units in the section be set to 0.  Since
the only $B$ and $C$ units in the entire section that contain a 1
are in the $N+1$st unit, and since the information in the $A$'s,
$B$'s and $C$'s can be shifted relative to eachother at will, the
1's in the $B$ and $C$ can be used as pointers to move bits around,
and to locate triples of bits on which one can operate with Fredkin
gates.

For example, to interchange $x_i$ and $x_j$, simply shift the
information in the $C$'s $N+1-i$ triples to the left.  The only
triple that has $C=1$ is now the $i$th triple.  Now act on all
triples $ABC$ with a
Fredkin gate with $C$ as the control input that determines whether
the other two inputs are to be interchanged.  Since the $i$th triple
is the only one in which $C=1$, this is the only triple in which
anything happens.  
In this triple, $x_i$ is moved from $A$ to $B$.
Now shift the information in the $B$'s and $C$'s $j-i$ triples to
the right.  The $j$th triple now contains $ABC= x_jx_i1$.  Act with
a Fredkin gate with $C$ as the control input, as before.  Once
again, there is only one triple in which $C=1$: the $j$th triple,
which after the operation of the gate takes the values
$ABC=x_ix_j1$.  Now shift the information in the $B$'s and $C$'s
$j-i$ triples to the left and operate with a Fredkin gate again.  The
$i$th triple now contains $ABC=x_j01$.  Shifting the information in
the $C$'s $N+1-i$ triples to the right results in the initial state,
but with $x_i$ and $x_j$ interchanged.

The {\it modus operandi} is clear: the $N$ bits in the $A$'s are the
sheep, and the two 1's in the $B$ and the $C$ are the shepherds.  By
the method of the previous paragraph, one can move to adjacent
triples groups of three bits on which one desires to act with a
Fredkin gate.  To operate with a Fredkin gate on the three bits $x,
y,z$ in
the $i$th, $i+1$st and $i+2$nd $A$'s, one performs the following
sequence of operations.  First, shift the information in the $C$'s
$N-i$ units to the left, and the information in the $B$'s $N-i-1$
units to the left.  The three triples now read $x00~y01~z10$:  
the $i+1$st triple is the only one that has
$C=1$, and the $i+2$nd triple is the only one that has $B=1$.
Operate on all triples with a Fredkin gate with $C$ as the control
input: the only triple affected is the $i+1$st, which now has
$ABC=0y1$.  Shift the information in the $C$'s $2$ units to the
right: the $C=1$ unit is now in the $i+3$rd triple, one
to the right of the three
triples under consideration, which 
read $x00~0y0~z10$.  Now operate on
all triples with a Fredkin gate with $B$ as the control input: the
only unit affected is the $i+2$nd, which now has $ABC=01z$.  Shift
the $B$'s to the left by one triple, and the $C$'s to the left by
two triples.  The three triples now read $xyz~011~000$.  Now operate
on all triples with a Fredkin gate with $A$ as the control.  The
only triple that can be affected is the $i$th: all other triples
that have $A=1$ have $B=C=0$.  The new values of $x,y,z$ can now be
moved back to their original positions simply by undoing the
reversible set of operations that brought them together.

By the above operations, one can move bits anywhere in the section,
act with Fredkin gates on any three bits that one desires, move bits
again, act on three more bits, etc.  This
method allows one to translate the
design for any logic circuit composed of Fredkin gates into a
sequence of pulses that realizes that logic circuit on the bits of
information within a section.  In each section, the same logical
circuit is realized.  After the operation of the circuit has been
completed, information can be exchanged between sections very
simply.  First, identify the bits that need to be transferred to the
section on the right.  By moving the $C$ control unit to the triples
in which those bits reside, each of those bits can be transferred
from the $A$ unit to the $B$ unit.  Now move the information in the
$B$'s $2N$ triples to the right, and transfer the information in the
desired bits from the $B$'s back to the $A$'s.  The transfer of
information from one section to the next is complete.  An analogous
procedure allows the transfer of information to the section on the
left.  The total number of pulses required to realize a particular
ciruit design is proportional both to the number of gates and the length
of the wires in the design.

The following is an even simpler method for inducing parallel
computing.  Its drawback is that it takes up more space and
requires more pulses to realize than the previous method.

Suppose that as before one has the circuit design for
the processors in the parallel computer, and that each processor
takes $N$ bits of input.  Let $m=\lceil N/3 \rceil$, the smallest
integer greater than $N/3$.  When loading information onto the
polymer, place the first $m$ bits of information on the $A$'s in the
section in which the processor is to operate, at intervals of $m$,
so that the first $A$ in the section contains the first bit, the
$m+1$th $A$ contains the second bit, the $2m+1$th bit contains the
third bit, etc., all other $A$'s in the section taking the value 0.
Now place the next $m$ bits on the $B$'s at intervals of $m+1$, and
the remaining bits on the $C$'s at intervals of $m+2$.  It can be
seen immediately that in each section, there is at most one triple
$ABC$ in which information is stored in adjacent units, and that
this is the only triple $ABC$ in which there can be more than a
single unit that takes the value $1$.  If the pulses that
effect a Fredkin gate are applied to the polymer, then this triple
is the only triple of bits in each section whose values can change,
since a Fredkin gate changes the values of its inputs only if more
than one of those bits takes the value $1$.  For each Fredkin gate
in the circuit diagram for  
the processor, the exchange operations
described above can be used to bring together in the same triple the
proper inputs to the gate, and the pulses that effect a Fredkin
gate can then be applied.  Any reversible logic circuit on the $N$
input bits can be realized in this fashion.  The exchange operations
can then be used to transmit information to the neighbouring
processors.  The total number of pulses required to enact the
circuit is proportional to the number of Fredkin gates, and
length of the wires, measured in terms of the
number of units over which information must be moved.

Both of the methods for inducing parallel computation described
above are easily adapted to providing parallel computation in more
than one dimension.

\bigskip
\noindent{\bf Quantum computation}

The resulting computer is not only a universal digital computer, but
a true quantum computer.  Bits can be placed in superpositions of 0
and 1 by the simple expedient of applying pulses at the proper
resonant frequencies, but of length different from that required to
fully switch the bit.  For example, if in loading information on the
polymer, as in the section above, instead of applying a $\pi$ pulse,
one applies a $\pi/2$ pulse of
frequency $\omega^{A:end}_0$ of length $T_1$, the effect is to
put the $A$ unit on the end in the
state, $1/{\sqrt 2}\big(|0\rangle + e^{-i\phi_1} |1\rangle
\big)$, where $\phi_1= \pi/2 + \omega^{A:end}_0 T_1$.  
Applied at a time $T_2$ later,
a $\pi$ pulse of frequency $\omega^B_{10}$ and length $T_3$
then puts the
first two units in the state, $1/{\sqrt 2} \big( |00\rangle +
e^{-i\phi_2}|11\rangle \big)$, where $\phi_2 = 3\pi/2 + \omega^{A:end}_0(T_1
+T_2) +
(\omega^{A:end}_1 + \omega^B_{10} )T_3$.

In fact, by the proper sequence of pulses, it is possible not only
to create any quantum state of $N$ bits, but to effect any
unitary transformation desired on those $N$ bits.  
The proof is by induction:  
The inductive assumption
is that it is possible to perform any unitary transformation on the
space spanned by vectors $|1\rangle,\ldots,|k\rangle$, where each
$|i\rangle$ is a member of the set, $\{ |00\ldots0\rangle,
|10\ldots0\rangle, \ldots, |11\ldots1\rangle \}$.  This is clearly
true for $k=2$, since it is possible by applying a resonant pulse of
the proper intensity and length to effect any desired unitary
transformation between the states $|000\ldots0\rangle$ and 
$|100\ldots0\rangle$, and since
it is possible by
the logical operations described above to arrange a sequence of
pulses whose only effect is to produce some desired
permutation of the states $\{ |00\ldots0\rangle, |10\ldots0\rangle,\ldots,
|11\ldots1\rangle \}$.\footnote{To derive this result, we use
methods developed by Bennett.$^5$
The desired permutation,$\Pi$ can be accomplished
by some reversible logical circuit, that takes as input the state $|i\rangle$,
and gives as output the state $e^{i\phi}|\Pi(i), j(i)\rangle$, where $j(i)$
is some `junk'
information that tells how the computation was performed,$^5$ and
$\phi$ is a phase whose value can be manipulated arbitrarily by
varying the length and intensity of the $\pi$ pulses used to effect
the computation.  The pulses can always be delivered in such a way
that $\phi=0$, the value which we will assume from this point on.  The
`junk' can be cleaned up by making a copy of $\Pi(i)$ and undoing the
computation, resulting in the state $|i, \Pi(i)\rangle$.   A second
circuit can perform the inverse transformation, $\Pi^-1(i)$, giving
the state $|i,i,j'(i)\rangle$, where $j'(i)$ is some more `junk.'
One of the two copies of $i$ can now be used to reversibly `uncopy'
the other,$^5$, leaving $|i,j'(i)\rangle$, and the inverse transformation
can be undone, leaving $|\Pi(i)\rangle$.}  
Now assume that the inductive assumption is true for some $k=n$: we
show that it must be true for $k=n+1$.  It suffices to show that one
can effect a unitary operation $U$ 
on the space spanned by $\{ |1\rangle, \ldots, |n+1\rangle \}$
that transforms an
arbitrary orthonormal
basis, $\{ |\psi_1\rangle, \ldots, |\psi_{n+1}\rangle \}$,
for that space so that $U|\psi_i\rangle=|i\rangle$.

Let $|\psi_{n+1}\rangle= \sum_{i=1}^{n+1} \alpha_i|i\rangle$.  By
the inductive hypothesis, one
can effect a unitary transformation that leaves $|n+1\rangle$ fixed
(over the course of time $t$, $|n+1\rangle$ acquires a multiplicative
phase,
$e^{-iE_{n+1}t/\hbar}$, but by varying the time over which the
pulses that effect the transformation are applied, this phase can be
made to take on an arbitrary value, which we will take to be 1), 
and takes $\sum_{i=1}^{n} \alpha_i|i\rangle$ to $\beta|0\rangle$:
the result is to take $|\psi_{n+1}\rangle$ to $|\psi_{n+1}\rangle' =
\beta|0\rangle+\alpha_{n+1}|n+1\rangle$.  Now one can effect a
unitary transformation that takes 
$\beta|0\rangle+\alpha_{n+1}|n+1\rangle$ to $|\psi_{n+1}\rangle'' =
|n+1\rangle$.  These transformations take the remaining
$|\psi_j\rangle$ to an orthonormal basis $\{ |psi_j\rangle'' \}$ for
the space spanned by
$|1\rangle,\ldots,|n\rangle$.  By the inductive hypothesis, 
one can effect a unitary transformation that leaves $|n+1\rangle$
fixed (once again, up to a phase that can be taken to be 1), 
and takes $|\psi_j\rangle''$ to $|j\rangle$.  The mapping is
complete, and the resulting proof shows not only that arbitrary
unitary maps can be constructed, but how to construct them.

The
proposed device is not only a universal digital computer, but a
universal quantum analog computer in the sense of Deutsch.$^8$  The
computer can be used to create and manipulate states that exhibit purely
quantum--mechanical features, such as Einstein--Podolsky--Rosen
correlations that violate Bell's inequalities.$^{16-18}$  In
addition, by giving each bit a quantum `twist' when loading it on the
computer (for example, by applying a $\pi/2$ pulse or a $3\pi/2$ pulse at
random), information could be encoded and stored in such a way that only the
person who knows by how much each bit has been rotated could read the
information.  All others who try to read it will get no information,
and will leave a signature of their attempt to read it in the
process, by
randomizing the states of the bits.$^{19}$ 

\bigskip
\noindent{\bf Dissipation and error correction}

Errors in switching and storing bits are inevitable.
It is clear that without
a method for error correction, the computer described here will not
function.

Error correction is a logically irreversible process, and requires
dissipation if errors are not to accumulate.$^3$
If in addition to a long--lived excited state, any of the
units possesses an excited state that decays quickly to a
long--lived state, this fast decay can be exploited to provide error
correction.  For example, each $B$ could have an additional excited
state, 2, that decays to the ground state, 0, in an amount of time
short compared with the time in between pulses.  Any $B$ in a long--lived 
state, 
1, e.g., can be restored to the ground state conditioned on the
state of its neighbours by applying pulses
with the resonant frequency $\omega_{ij}^B(12)$ of the transition
between the states 1 and 2 given that its neighbours $A$ and $C$ are
in the states $i$ and $j$.  The pulse need not have a well--defined
length, provided that it is long enough to drive the transition
efficiently.
   
If just one type of unit has
a fast decay of the
sort described, then one can realize not only any reversible
cellular automaton rule that updates first one type of unit, then
another, but any irreversible cellular automaton rule
as well.  The scheme described above that allows the construction of
one--dimensional arrays of arbitrary parallel--processing reversible
microprocessors then allows one to produce one--dimensional arrays
of arbitrary irreversible microprocessors, each one of which can contain
arbitrary error--correcting circuitry.  
Many error--correcting 
schemes are possible, using check sums and parity bits$^{20-21}$
multiplexing,$^{22}$ etc.  A particularly simple and robust scheme is given
below.  
For each logically
irreversible operation accomplished, a photon is emitted
incoherently to the environment.  In contrast to the switching of
bits using $\pi$ pulses, in which photons are emitted and absorbed
coherently, the switching of bits using fast decays is inherently
dissipative.  The amount of dissipation depends on what is done with
the incoherently emitted photons.
If the photon is absorbed and its energy
thermalized, then considerably more than $k_BT$ is dissipated; if
the energy of the photon is put to work, dissipation can be brought
down to close to $k_BT$.

Such a computer can 
function reliably in the face of a small error rate in principle.
Error correction for the method of computation proposed here takes
the place of gain and signal restoration in conventional circuits.
Whether such a computer can
actually be made to function reliably in the face of a finite error
rate 
depends crucially on whether the error correction routine suffices
to correct the number of errors generated in the
course of the computational cycle, in between error correction
cycles.

Suppose that the probability of error per unit per computational cycle is
$\epsilon$.  Suppose that all bits come in $2k+1$ redundant copies,
and that after each cycle, error correction is performed in parallel
by having the copies vote amongst eachother as to their proper
value, and all copies are restored to that value: there exist quick
routines for performing this operation, that are insensitive to
errors generated during their execution.  The error rate
per cycle is reduced to $\eta \approx (2\epsilon)^k$ by this
process.  For a computation that uses $b$ bits over $c$ cycles to
have a
probability no greater than $f$ for the 
failure of a single bit, we must have $b(1-\eta)^{bc} \geq  b-f $,
which implies that $\eta\leq 1/cb^2$.  For example, suppose that the
error rate per bit per cycle is a quarter of a percent, $\epsilon=.0025$.  To
have a computation involving $10^{12}$ bits over $10^{20}$ steps
have a probability of less than $1\%$ of getting a bit wrong
requires that each bit have $47$ redundant copies.  Although such
computers have much higher error rates and require much more error
correction than conventional computers, because of their high bit density
and massively parallel operation, error correction can
be carried out without too great a sacrifice in space or time.  

\bigskip
\noindent{\it Robust error correction schemes}

As noted, a wide variety of error correction schemes are possible.
Here we present several schemes that are robust: they 
correct errors quickly and efficiently, 
even if errors are committed during their
execution.

First, we examine the simplest possible form of error correction.
If information is stored redundantly in triplicate, so that each
$ABC$ is supposed to contain the same bit of information, a
simple form of error correction is provided by applying a sequence
of pulses that restores $ABC$ to $111$ if at least two of the units
are $1$, and to $000$ if at least two of the units are $0$.

Suppose that each $B$ unit has an excited state, call
it 2, that decays to the ground state 0 in an amount of time
comparable to the length of the $\pi$ pulses used to do switching.
To restore $B$ to 0 if $A$ and $C$ are both 0, one simply applies a
pulse at the resonant frequency $\omega^B_{00}(12)$
of $B$'s transition from 1 to 2,
given that $A=0, C=0$.  As noted above,
the pulse need not be of any specific length, as long as it is
considerably longer than the lifetime of the state 2.  This pulse
has the following effect:  If $B$ is
initially in the state 0, it remains in that state.  If $B$ is
initially in the state 1, then it is excited to the state 2, and
then decays to 0.  The net effect is to reset $B$ to 0 provided that $A$
and $C$ are both 0.  To restore $B$ to 1 if both $A$ and $C$ are 1,
first apply a pulse at the resonant frequency
$\omega^B_{11}(12)$ of $B$'s transition from 1 to 2,
given that $A=1$, $C=1$.  This pulse insures that if $A$ and $C$ are
1, $B$ is set to 0.  Then apply a $\pi$ pulse of frequency
$\omega^B_{11}(01)$ to take $B$ from 0 to 1.  The net
effect is to reset $B$ to 1 provided both $A$ and $C$ are 1.
If $A$ and $C$ have different values, the sequence of pulses 
has no effect on $B$.  A simple sequence of
pulses will interchange the bits in $B$ and $C$.  If one
interchanges $B$ with $C$, and performs the resetting procedure of
the previous paragraph, then interchanges $A$ with $B$ and resets
$B$ once again, the required error correction is accomplished.

What is desired is a method of correcting errors by the method of
voting, but to have the voting take place over an arbitrary number
of copies.  One way to do this would be to design a circuit that
performs the voting, and then realize it by the method given above.
Such a circuit might take a large number of pulses to realize,
however, and would moreover be vulnerable to errors committed in the
course of its operation.  We have designed some quick and dirty
methods that perform error correction massively in parallel.  The
basic idea is to store all bits with $n$-fold redundancy, in blocks,
to have each bit in the block vote with its neighbours in
threes, as
above, then to scramble up the bits in the block and to perform the
voting again.  If the error rate per computational step is
$\epsilon$, and if no errors are made in the voting, 
then after a single vote, the rate will be $\epsilon^2$,
after two votes, $\epsilon^4$, etc.:  the
the effective error rate rapidly drops to zero.  If during voting, there is a
probability of error per unit of $\theta$, then the fraction of units in
each block with the `wrong' value eventually converges to
$\theta$.  The process described is insensitive to errors
committed during its execution.

There are several ways to realize this particular method of error
correction.  In the methods for inducing computing
described above, we have some bits stored in the $A$'s, some in the
$B$'s, and some in the $C$'s:  Suppose that each bit is stored with
$n$-fold redundancy, with $n$ triples of blank space between each
reduntantly registered bit.  When error correction begins, then, the
data is stored in blocks of $n$ triples, and in each block, all the
$A$'s are supposed to be the same, with all the $B$'s and $C$'s
equal to zero, or all the $B$'s are supposed to be the same, with
all the $A$'s and $C$'s equal to zero, etc.  Between the blocks of
data, there are blocks $n$ triples long in which all units are 0.
Both of the computational methods described above can be made to work with
this formatting of the data.

The error correction routine restores first the $A$'s to a common
value, then the $B$'s, then the $C$'s.  The method is simple: To
restore the $A$'s, first supply a sequence of pulses that transforms
$ABC=100$ to $ABC=111$, and $ABC=111$ to $ABC=100$,
leaving all other values for $ABC$ unchanged.  Such a
sequence is easy to devise.  This transformation has the effect
of making each unit in the block in which information is stored in
$A$'s take on the value of the $A$ in the triple, while leaving the
blocks in which information is stored in $B$'s or $C$'s unchanged.  
One can now make the different $A$'s in the block 
vote three by three amongst themselves.
First, 
shift the information in the $B$'s in one direction by some number of
triples $\leq n$, and shift the $C$'s to the opposite direction by some
independently chosen number $\leq n$; then restore an $A$ to one
if its neighbours are one, and to zero if its neighbours are zero.
By differing the shifts, one can make each $A$ in the block vote with
any $A$ to its left and any $A$ to its right.  
After the voting is done once, one can shift the $B$'s and the $C$'s back 
again, and perform
the inverse transformation within each triple, mapping $111$ to
$100$, and $001$ to $111$,
leaving other values fixed.  

The only weakness in this voting scheme
occurs at the ends of blocks in which all the $A$'s are supposed to
be equal to 1: the last $A$ on each block
doesn't get to vote.  As a result, without some further measure,
blocks in which all the $A$'s are supposed to be equal to 1 will
tend to be eaten away from the ends.   
This problem is easily remedied by scrambling up the redundant bits
in each block.  There is a wide variety of ways to induce this scrambling, 
one of the simplest of which is the following.

The
redundant bits in blocks where information is stored in the $A$'s
can be scrambled amongst
eachother by inducing the following interaction between that block
and a block in which the $C$'s are supposed to be equal to 1, (for
example, the block of $C=1$ that is used to shepherd bits around in the first
computing scheme above).  
First, shift the information in the blocks relative to eachother so
that the $m$ triples on the right of the block in which information
is stored in the $C$'s overlaps the first $m$ triples on the left of
the block in which information is stored in the $A$'s,  
where $m\leq n/2$.  Enact a Fredkin gate with $C$ as the
control bit.  Since the $C$'s are by and large equal to 1, the
effect is to interchange the information in the $m$ $A$'s of
the overlapping blocks with the 0's in the $m$ $B$'s. 
Shift the
information in the $B$'s and $C$'s $m$ triples to the right, and act
with a Fredkin gate with control $C$ again.  Then shift the
information in the $B$'s and the $C$'s $m$ triples to the left, and
act with a Fredkin gate with control $C$ again.  
The effect of these
actions is as follows:
the only place in which anything happens is in the
first $2m$ triples of the blocks where information is stored in the
$A$'s,
as in all other triples where $C=1$, $A=B=0$.  
In such blocks, the effect is to interchange
the information that was in the first $m$ $A$'s of the blocks 
with the information in the second $m$ $A$'s of the blocks.
Similarly, it is possible to interchange the information in the last
$m$ $A$'s of the blocks with the information in the second to last
$m$ $A$'s of the blocks.  By varying $m$, one can scramble up the
$A$'s in any way that one chooses.  Similarly, one can use a block
of $A=1$ to scramble up the bits in blocks in which information is
stored in $B$'s or $C$'s.  There are other, more involved scrambling
procedures that do not require blocks of $C=1$ to scramble up the
bits in blocks in which information is stored in the $B$'s or
$C$'s.

Combining such a scrambling procedure with the procedure of voting by
threes is an effective method for correcting errors quickly as long as
each bit has a sufficiently large number of redundant copies that the
probability of more than a third of them taking on the wrong value in
the course of the computation is small.
If the probability of error
within a block of $A$'s after the computational cycle and before the
error correction cycle was $\epsilon$, 
and if the
probability of error per unit during the error--correction cycle is
$\theta$,  then the probability of error after one round of voting
by threes is $\approx 2\epsilon^2+\theta$.  More precisely,
if the number of
incorrect $A$'s in the block was initially $p$, 
the probability that an $A$ in the block takes on the
correct value after voting once by threes is $\approx
1-p^2/n^2(2-p/n)-\theta$.
One can now
repeat the process, having each $A$ vote with the copies of a
different
pair of $A$'s in the same block.  The probability of error is now
reduced to $\approx 2(2\epsilon^2+\theta)^2+\theta$.  Etc.  As long
as the number of redundant bits is sufficiently large, this method
will rapidly restore all but a fraction $\theta$ of the bits in a
block to a common value.  Here, `sufficiently large,' depends on how
many times the voting by triples takes place.  Because it takes
place in parallel fashion, each voting requires only a short
sequence of pulses to realize.  For only a small number of votes,
five, say, it suffices that the  
number of incorrect $A$'s in the block, $p$, never gets larger
than $n/3$ in the course of the computation.

The above method, if error--free, leaves unaffected blocks in which
only the $B$'s or only the $C$'s contain data.  If the error rate is
$\theta$ over a single vote and scramble 
of the $A$'s, then the error--correcting
routine for the $A$'s will introduce an error rate of
$\approx\ell\theta$ per unit of the $B$'s and $C$'s, if carried out
over $\ell$ votes.  The same method can now
be used to restore first the $B$'s, then the $C$'s to their proper values.
The resulting error correction technique is efficient and robust.
It can easily be generalized to higher dimensions, and to more types of
units with more states.

\bigskip
\noindent{\it Destruction of quantum coherence}

Note that when a photon is emitted incoherently, the quantum
coherence of the bit from which it was emitted, and of any other bits
correlated with that bit, is destroyed.  Incoherent processes and
the generation of errors intrinsically limit the number of steps
over which the computer can function in a purely quantum--mechanical
fashion.

\bigskip
\noindent{\bf Errors}

There are many potential sources of error in the
operation of these pulsed quantum computers.  The primary difficulty
in the proposed scheme is the delivery of effective $\pi$ pulses.
Microwave technology can give complete inversion with error rates of
a fraction of a percent in NMR systems.  Optical systems are at
present harder to invert, since bands suffer considerable
homogeneous and inhomogeneous broadening.
As noted above, a fraction
of a percent error per bit per pulse
can be tolerated; but a few percent is probably
too much.  Techniques such as pulse shaping$^{23}$ or iterative
excitation schemes$^{24}$ enhance $\pi$ pulse effectiveness and
selectivity.
If optical systems with sufficiently narrow bands can be found, and
if the systems can be well--oriented, so that the coupling with the
pulses is uniform, then 
the rapid advance of laser technology promises soon to 
reach a level at which
$\pi$ pulses can be delivered at optical
frequencies. 

In addition to the technological problem of supplying accurate $\pi$
pulses, the following fundamental physical effects can cause
substantial errors:

\bigskip
\noindent{\it Effect of off--diagonal terms in interaction
Hamiltonians.}  These terms have a number of effects.  The simplest
is to induce unwanted switching of individual units, with a
probability of error per unit per pulse
of $\big( \delta\omega_{off}/\omega \big)^2$ whenever a unit or its
neighbour is switched.
Here
$\hbar\delta\omega_{off}$ is the characteristic
size of the relevant off--diagonal
term in the interaction Hamiltonian.  Off--diagonal interactions
also induce the propagation of excitons along the polymer: this
process implies that a localized excited state has an intrinsic
finite lifetime equal to the inverse of the bandwidth for the
propagation of the exciton associated with that state.$^{25-26}$  For the
polymer $ABCABC\ldots$, the bandwidth associated with the
propagation of an excited state of $A$ can be calculated either by a
decomposition in terms of Bloch states, or by perturbation theory,
and is proportional to
$\delta\omega^{AB}_{off}\delta\omega^{BC}_{off}\delta\omega^{CA}_{off}
/ (\omega_A-\omega_B)(\omega_A-\omega_C)$, where $\hbar\delta 
\omega^{AB}_{off}$, e.g., is the size of the term in $H_{AB}$ that induces
propagation of excitation from $A$ to $B$.  For a polymer of the
form $12\ldots M 12\ldots M \ldots$, the characteristic bandwidth
goes as $\delta\omega_{off}^M/\Delta \omega^{M-1}$, where $\Delta
\Omega$ is the typical size of the difference between the resonant
frequencies of different types of units.  For the computer to
function, the inverse of the exciton propagation bandwidth must be
much longer than the characteristic switching time.  If the
off--diagonal terms are of the same size as the on--diagonal terms,
on average, then for the computer to function, the overall
interaction between units must be weak, and 
$M$, the number of different kinds of units in the
polymer, must be at least three.  Small off-diagonal terms and a
relatively large number of different types of units are essential
for the successful operation of the computer.

\bigskip
\noindent{\it Quantum--electrodynamic effects.}  The probability of
spontaneous emission from a single unit is assumed to be small.  In
the absence of interactions, the spontaneous decay rate for a unit
with resonant frequency $\omega$ is $4\omega^3\mu^2/3\hbar c^3$.$^{13}$  If
the lifetime of an optical excited state is to be as long as milliseconds,
the induced dipole moment $\mu$ must be suppressed by symmetry
considerations.  
Interactions between different units of the same type can give rise
to quantum--electrodynamic effects such as super--radiance, and the
coherent emission of a photon by one unit and coherent reabsorption
by another.$^{27}$
Fortunately, the states that are being used for
computation, in which each unit is in a well--defined excited or
ground state, are exactly those that
do not give enhanced probabilities for these
processes.  In the process of switching, however, and when bits are
in superpositions of $|0\rangle$ and $|1\rangle$, super--radiant
emission gives an enhancement of the spontaneous emission rate by a
factor of $n$,
where $n$ is the number of units of the same type
within a wavelength of the light used.  Since the switching time is
short compared to the lifetime, super--radiant emission is not a
problem. (Though super--radiance
can shorten the lifetime of quantum superpositions of logical
states.) 

\bigskip\noindent{\it Nonlocal interactions.}  
Coherent switching
will not work unless the shift in a unit's resonant frequency
induced by nearby units of the same type is too small to throw the
unit out of resonance.  
The dipole--dipole
couplings of reference 2 fall off as $1/r^3$.  For such a long
range coupling, many different types of units are
required, and the result of a resonant pulse is to realize a
cellular automaton rule with a neighbourhood of radius larger than
one.  

\bigskip
None of the purely physical effects gives error rates that are
insurmountable.  If the $\pi$ pulses 
are long compared to the
inverse frequency shifts due to interaction, if the unperturbed
resonant frequencies differ substantially between the different
types of unit, and if the off--diagonal
terms in the interaction Hamiltonians are small compared with the
resonant frequencies and their differences, then this computing
scheme will work in principle.    

Although putting them together in a working
package may prove difficult,  
precisely timed monochromatic
laser pulses, well--oriented polymers, accurately fabricated
semi--conductor arrays, and fast, sensitive photodetectors are all
available in today's technology.  Continuously tunable ti-sapphire
lasers, or diode--pumped YAG lasers tuned by side--band modulation
can currently supply frequency--stable picosecond pulses at nanosecond
intervals 
with an integrated intensity that varies by
a fraction of a percent.  Currently available 
electro--optical shutters could be used to
generate the proper pulse sequence at a nanosecond clock rate.
Photodetectors equipped with
photomultipliers and acoustic--optical filters can reliably detect
tens of photons (or fewer) within a wavelength band a few nanometers
wide.  Although arrays of quantum dots created by $X$-ray
lithography are not yet of sufficiently uniform quality,
arrays of quantum dots and
lines that have been created using interference techniques might
be sufficiently uniform to realize the proposed scheme.

\bigskip
\noindent{\bf Numbers}

The range of speed of operation of such a pulsed 
quantum computer within acceptable error rates is determined by the
frequency of light used to drive transitions, and by the strength
and character of
the interactions between units.  For square-wave pulses, 
the intrinsic probability of error  per unit per pulse
due to indiscriminate transition--driving
is  $\big(1/T\delta\omega_{on}\big)^2$, where $T$ is the pulse length
and $\delta\omega_{on}$ is resonant frequency shift induced by on-diagonal
terms in the interaction Hamiltonian,$^{13}$ (this error can be reduced
significantly by using shaped pulses$^{26}$) while the probability of
error per unit per pulse due to off--diagonal terms in the
interaction Hamiltonians is $\big( \delta\omega_{off}/ \omega
\big)^2$.  The decay of localized excitations due to exciton
propagation gives a lifetime proportional to $\Delta \omega^{M-1}/\delta
\omega_{off}^M$, where $M$ is the number of different types of
units.  

Suppose that
the excited states have transition frequencies corresponding
to light in the
visible range, say $\omega=10^{15}$ sec$^{-1}$.  (Many electronic
excited states in molecular systems and quantum dots are in the
visible or near--visible
range.  Visible light is a good range in which to operate, because
accurate lasers exist for these frequencies, and because the systems
can operate at room temperature.)  In the absence of off--diagonal
terms in the interaction Hamiltonians, the frequency shifts due to
interaction do not
need to be small compared to $\omega$, and to obtain an intrinsic
error rate of less than $10^{-6}$ per unit per pulse, the pulse
length could be as short as $10^{-12}$ seconds, and as long as a few
thousands of the intrinsic lifetimes of the excited states (assuming
that a few thousand pulses are required for error correction).  The
clock rate of such a computer could be varied to synchronize its
input and output with conventional electronic devices.
In
the presence of off--diagonal terms of the same magnitude 
$\delta\omega_{off} \approx \delta\omega_{on} \sim \delta\omega$ as the
on--diagonal terms, to obtain an intrinsic error rate of $10^{-6}$
per unit per pulse, one must have $\delta \omega = 10^{12}$
sec$^{-1}$, and a minimum pulse length of $10^{-9}$ seconds.   If
the computer has three different types of units, the intrinsic
exciton lifetime from the local coupling alone
is on the order of $10^{-6}$ seconds.  
Actual exciton lifetime will be shorter as a
result of coupling to other modes.  
The more different kinds of units, the more
freedom one has to lengthen the clock cycle.

If the units in the quantum computer are nuclear spins in an intense
magnetic field, with dipole--dipole interactions,
then the the pulses will have frequencies in the
microwave or radiofrequency region, and the computers will have
clock rates from microseconds to milliseconds.    

\bigskip
\noindent{\bf Conclusion}

Computers composed of arrays of pulsed, weakly--coupled quantum
systems are physically feasible, and may be realizable with current
technology.  The units in the array could be quantum dots, nuclear
spins, localized electronic states in a polymer, 
or any multistate quantum system
that can be made to interact locally with its neighbours, and can be
compelled to switch between states using resonant pulses of light.
The exposition here has concentrated on one-dimensional 
systems with two or three
states, but more dimensions, more types of unit, and more
states per unit provide higher densities of
information storage and a wider range of possibilities for
information processing, as long as the different
transitions can still be driven discriminately.

The small size, high clock speeds and massively parallel operation
of these pulsed quantum computers, if realized, would
make them good devices for simulating large, homogeneous systems
such as lattice gases or fluid flows.  
But such systems are capable of more than digital computation.  
When operated coherently, the
devices described here are true quantum computers, combining digital
and quantum analog capacities, and could be used
to create and manipulate complicated many--bit quantum states.  
Many questions remain: What are the best physical
realizations of such systems?  (The answer may be different
according to whether the devices are to be used for fast, parallel
computing, or for generating novel quantum states.)  
How can they best be programmed?  How can noise be suppressed and
errors corrected?  How can their peculiarly quantum
features be exploited? 
What
are the properties of higher dimensional arrays?  The device
proposed here, as with all devices in the next
generation of nanoscale information processing,
cannot be built and made to function without addressing fundamental
questions in the physics of computation.

\vfill
\noindent{\bf Acknowledgements:} The author would like to thank
G\"unter Mahler, Carlton Caves, Alan Lapedes, 
Ronald Manieri, and Brosl Hasslacher
for helpful discussions.
\eject

\centerline{\bf References}
\bigskip
\item{1.} S. Lloyd, {\it A potentially realizable quantum computer},
submitted to {\it Science}.

\item{2.} K. Obermayer, W.G. Teich, G. Mahler, {\it Phys. Rev.} {\bf
B 37}, 8096--8110 (1988).  W.G. Teich, K. Obermayer, G. Mahler, {\it
Phys. Rev.} {\bf B 37}, 8111--8121 (1988).
W.G. Teich, G. Mahler, {\it Phys. Rev.}
{\bf A 45} 3300, 1992.

\item{3.} R. Landauer, {\it IBM J. Res. Develop.} {\bf 5},
183--191 (1961).

\item{4.} K.K. Likharev, {\it Int. J. Theor. Phys.} {\bf 21},
311-326
(1982).

\item{5.} C.H. Bennett, {\it IBM J. Res. Develop.} {\bf 17},
525--532 (1973); {\it Int. J. Theor. Phys.} {\bf 21},
905--940 (1982).

\item{6.} E. Fredkin, T. Toffoli, {\it Int. J. Theor. Phys.} {\bf
21}, 219-253 (1982).

\item{7.} P. Benioff, {\it J. Stat. Phys.} {\bf 22}, 563--591
(1980); {\it Phys. Rev. Lett.} {\bf 48}, 1581--1585
(1982); {\it J. Stat. Phys.} {\bf 29}, 515--546 (1982); {\it Ann.
N.Y. Acad. Sci.} {\bf 480}, 475--486 (1986).

\item{8.} D. Deutsch, {\it Proc. Roy. Soc. Lond.} {\bf A 400},
97--117 (1985); {\it Proc. Roy. Soc. Lond.} {\bf A 425}, 73--90
(1989).

\item{9.} R.P. Feynman, {\it Optics News} {\bf 11}, 11-20 (1985);
{\it Found. Phys.} {\bf 16}, 507--531 (1986); {\it Int. J. Theor.
Phys.} {\bf 21}, 467--488 (1982).

\item{10.} W.H. Zurek, {\it Phys. Rev. Lett.} {\bf 53}, 391--394
(1984).

\item{11.} A. Peres, {\it Phys. Rev.} {\bf A 32}, 3266--3276 (1985).

\item{12.} N. Margolus, {\it Ann. N.Y. Acad. Sci.} {\bf 480},
487--497 (1986); {\it Complexity, Entropy, and the Physics of
Information, Santa Fe Institute Studies in the Sciences of
Complexity} {\bf VIII}, W.H. Zurek, ed., 273-288 (1991).

\item{13.} L. Allen, J.H. Eberly, {\it Optical Resonance and
Two--Level Atoms}, Wiley, New York 1975.

\item{14.} W.H. Louisell, {\it Quantum Statistical Properties of
Radiation}, Wiley, New York 1973.

\item{15.} S. Wolfram, {\it Rev. Mod. Phys.} {\bf 55}, 601 (1983).

\item{16.} A. Einstein, B. Podolsky, N. Rosen, {\it Phys. Rev.} {\bf
47}, 777 (1935).

\item{17.} G. Mahler, J.P. Paz, to be published.

\item{18.} D.M. Greenberger, M. Horne, A. Zeilinger, in {\it Bell's
Theorem, Quantum Theory, and Conceptions of the Universe}, M.
Kafatos, ed., Kluwer, Dordrecht, 1989.  N.D. Mermin, {\it Phys.
Today} {\bf 43}(6), 9 (1990).

\item{19.} C.H. Bennett, G. Brassard, S. Breidbart, S. Wiesner,
{\it Advances in Cryptology: Proceedings of Crypto `82}, Plenum, New
York, 267-275
(1982).  S. Wiesner, {\it Sigact News} {\bf 15}(1), 78-88 (1983).

\item{20.} C.E. Shannon, W. Weaver, {\it The Mathematical Theory 
of Information},
University of Illinois Press, Urbana 1949.

\item{21.} R.W. Hamming, {\it Coding and Information Theory}, Prentice--Hall,
Englewood Cliffs, 1986.

\item{22.} J. von Neumann, {\it Probabilistic Logics and the
Synthesis of Reliable Organisms from Unreliable Components},
Lectures delivered at the California Institute of Technology, 1952.   

\item{23.} W.S. Warren, {\it Science} {\bf 282}, 878-884 (1988).

\item{24.} R. Tycko, A. Pines, J. Guckenheimer, {\it J. Chem. Phys.}
{\bf 83}(6), 2775-2802 (1985).

\item{25.} R.S. Knox, {\it Theory of Excitons}, Academic Press, New
York 1963.

\item{26.} P.W. Anderson, {\it Concepts in Solids}, Addison--Wesley,
Redwood City 1963.

\item{27.} R.H. Dicke, {\it Phys. Rev.} {\bf 93}, 99 (1954).

\vfill\eject\end

\item{11.} F.L. Carter, {\it Molecular Electronics} and {\it
Molecular Electronics II}, Dekker, New York, 1982 and 1987.  

\item{12.} D.K.
Ferry, J.R. Barker, C. Jacobini, {\it Granular Nanoelectronics},
Plenum, New York, 1991.
K.E. Drexler, {\it Nanosystems}, Wiley, New York, 1992.
See also the now--defunct {\it Journal of Molecular Electronics},
and the proceedings of the {\it International Symposia of Molecular
Electronics}.

\item{13.} J.J. Hopfield, J.V. Onuchic, D.N. Beratan, {\it Science}
{\bf 241}, 817--820 (1988).

\centerline{\bf Appendix}

\vfill\eject
\item{(1)}  {\it Indiscriminate driving of transitions.}  
The
probability of a $\pi$ pulse 
with  frequency $\omega$ 
driving a transition with frequency $\omega'$ is equal
to $1/\big((\omega-\omega')^2 T^2+1\big)$, where $T$ is the duration
of the $\pi$ pulse.$^{15-16}$  To prevent indiscriminate driving of the wrong
transition, the resonant frequencies of the different units should
differ, and the pulse length should be much longer than the inverse
of the characteristic perturbation in frequency due to a change in
the state of a unit's neighbours.  To drive the transition
accurately, the frequency of the $\pi$ pulse must be centered on the
transition frequency to an accuracy greater than $1/T$.  

\item{(2)}  {\it Inefficient driving of transitions due to 
incorrect pulse length and height.}  
If the integrated intensity over the pulse length is off, so
that $\hbar^{-1} 
\int \vec \mu_B \cdot \hat e {\cal E}(t) dt = \pi+\delta$, then the
error goes as $sin^2(\delta)$.

\item{(3)}  {\it Nonlocal interactions.}  The dipole--dipole
interactions used in reference 3 fall off only as one over
distance cubed.
For driving transitions to a state with a rapid
decay, where pulse intensity and length need not be specified exactly, 
this can
be compensated for by having a pulse with a finite bandwidth.$^3$
For driving coherent transitions, this does not suffice.  To drive
coherent transitions effectively, one first needs a sufficient
number of different types of units so that the frequency shift of a
unit due
to units of the same type, which are being switched by the same
pulse, is negligible (otherwise, the frequency
shifts due to these units will change in the course of switching).
One must then give pulses that take into account the frequency shift
due to a unit's next nearest neighbours, next to next nearest
neighbours, etc.  
The resulting computer is a cellular automaton with a neighbourhood
of radius greater than one. 
The size of the
neighbourhood depends on how local the interaction is: one can
ignore the frequency shifts due to units sufficiently far away that
the error rate induced by such shifts is acceptible.
Different realizations of this computational scheme have different
advantages.  Polymers are easy to fabricate, but are
one--dimensional, and so more limited in the parallel computation
that they can perform (molecular crystals of polymers are a natural
way to produce a more complicated computational network).  Nuclear
spins are easy to arrange in lattices, and provide for three
dimensional computing, but must operate at lower energies, slower
speeds and lower temperatures, (in addition, one must develope ways
to compensate for the long--range nature of the dipole--dipole
interaction). 
Quantum
dots can be produced in two--dimensional arrays that need not be
periodic, in which complicated patterns of dots can be embedded at
will.  The arrays and the dots themselves need to be manufactured to
exacting tolerances in order for the resonance methods
detailed here to provide reliable switching.